\documentclass[
10pt]{article}
\usepackage{amssymb,amsfonts,latexsym,stmaryrd}
\usepackage{amsmath}
\usepackage{graphicx}

\renewcommand{\emph}[1]{\textbf{#1}}

\title{Robin Milner's Work on Concurrency:\\ An Appreciation}

\author{Samson Abramsky}
\date{MFPS 2010}

\begin{document}

\maketitle

\begin{abstract} 
We give a short appreciation of Robin Milner's seminal contributions to the theory of concurrency.
 \end{abstract}

\section{Introduction}


This brief note is based on the talk I was asked to give at a session honouring Robin's work which took place at MFPS in Ottawa, on May 8th 2010, just a few weeks after the sad news that he had passed away.

Bob Harper spoke in the same session on Robin's work on LCF and ML. I decided to focus on his work on concurrency. I believe that this was the work which meant the most to him --- and which was the deepest of his many remarkable achievements.

In 1989 I wrote\footnote{The context was a letter  in support of Robin's nomination for a Turing award, which he of course received in 1991.} that Robin's ideas had become part of the air we breathe in our scientific community.
That is even more true today!

There is so much we have learnt, and can continue to learn, from Robin, over and above technical details of this or that formalism.
I will try to articulate just a little of this in the brief remarks which follow.

\section{1979: The First Revolution}

Let us begin with some background.

The 1970's had seen an extensive development of domain theory and denotational semantics.
The basic paradigm here is that
programs are modelled as \emph{functions}. 
This is fine for functional programming  (\textit{pace} sequentiality);
less good for effects; 
but what about concurrency?

There was already a theory of concurrency, Petri's Net Theory. This contained some deep ideas, but was lacking some critical structural features. There were also a number of important programming ideas: Dijkstra's semaphores, Brinch Hansen's monitors, Hoare's critical regions etc., but no systematic and comprehensive theory.

At the same time, there were urgent technological imperatives: distributed systems, the Internet, etc.
Who would meet the challenge?

\subsection*{Enter CCS}

Robin's introduction of CCS transformed the field. 
The crucial ingredients seem to me the following:

\begin{itemize}
\item Processes are seen as a \emph{mathematical structure} (rather than a machine model). 
\item In particular, processes are given an \emph{algebraic structure}.
\[ P \;\; ::= \;\; a.P \mid \mathbf{0} \mid P+Q \mid P \parallel Q  \mid P\setminus a \mid P[S] \mid \ldots \]
This has been enormously important: for the first time, processes could be studied \emph{compositionally}.
\item We knew what functions were already! But what are processes?
CCS established the fundamental methodological point of studying them through their \emph{observational behaviour} (defined in terms of labelled transition systems via SOS).

This led inexorably in turn to notions of \emph{observational equivalence}.

\item In the train of these ideas came a host of technical developments: equational axiomatizations, Hennessy-Milner logic, algorithmics of bisimulation, and more.

\end{itemize}

\noindent CCS was not just a new calculus, but a new \emph{paradigm} --- that has played a central r\^ole in the subsequent development of our subject.
It opened up the world of compositional behavioural modelling.
%

\subsection*{MFPS 1989}

Many people would have rested on their laurels. Robin never did.

\begin{itemize}
\item Having revolutionized the field with his CCS book in 1980 \cite{milner1980calculus}, in large part stimulating the growth of a whole research community, including related developments such the extensive work on CSP (Brookes, Hoare, Roscoe et al.) and the Dutch school of process algebra (Bergstra, Klop et al.); 

\item Having led the effort to standardize ML and personally crafted, with Bob Harper, Mads Tofte et al.~the formal definition and commentary on the language \cite{milner1997definition,milner1991commentary}; 

\item  Having written the masterly text  \textit{Communication and Concurrency} \cite{milner1989communication} in 1989, as a polished and definitive presentation of CCS:

\end{itemize}

\begin{quotation}
``They might have said: `He stopped here'.''\footnote{Robin said this to me during MFPS 1989 --- a rare occasion when I heard him consider what might have been, rather than focus intensely on what was.}
\end{quotation}

\noindent This is the opposite of what happened. Robin's talk at MFPS was one of the first occasions where he presented what was to become known as the \emph{$\pi$-calculus}.

At that meeting he was re-energized, brimming with ideas. A whole new phase of the subject was about to begin!

\subsection*{Robin's quest: the $\lambda$-calculus of concurrency}

The $\lambda$-calculus is an incredibly rich, concise, comprehensive theory of functions and functional computation.There are infinitely many (fruitful) extensions and variations, but the core calculus can be presented in a few lines --- and there is (essentially) only one!

Could something of similar quality be achieved for concurrency?

For all the success of CCS and its variants, Robin realized --- well before anyone else (and some still haven't!) --- that they had not achieved this goal.
As one telling and key example, the $\lambda$-calculus itself cannot be compositionally encoded in CCS   in a satisfactory fashion.

With a remarkable leap of insight, Robin saw that one could use the notion of \emph{name} as a building block to allow the expression of \emph{mobility} of processes, and hence to open up a huge increase of expressive power in process calculi.

The importance of naming, unique and fresh names etc.~had long been known to systems researchers; but no-one had imagined how this could be distilled into such an elegant and expressive theory.

\section{The $\pi$-calculus}

The $\pi$-calculus developed from the initial version introduced by Robin with Joachim Parrow and David Walker \cite{milner1992calculus}, into a beautiful, concise, and expressive calculus. 

The core calculus can indeed be conveyed in a single line:
\[ P \;\; ::= \;\; x(y).P \mid \bar{x}y.P \mid \mathbf{0} \mid P\parallel Q \mid \nu x.P \mid \, !P \ldots \]
Again, what was created was not just a particular calculus, but a \emph{paradigm}.
It was to spawn whole sub-genres of `mobile' and `nominal' calculi.

Some key ingredients of the $\pi$-calculus:
\begin{itemize}
\item Structural congruence. This was Robin's transmutation of ideas from Berry and Boudol's Chemical Abstract Machine, and has proved an extremely useful idea which has been widely adopted.
\item Names, freshness, scope extrusion. This led to a whole sub-paradigm of `nominal calculi', in functional as well as process forms.
\item Behavioural types, which constrain dynamical process behaviour, rather than statically classifying values as in classical type theories. 
\item The ability to represent a huge range of computational phenomena, including higher-order and object-oriented computation, security protocols, biological modelling, business processes, and more.
\end{itemize}

\noindent The $\pi$-calculus proved to be even more hugely successful than CCS.
Robin gave another polished and masterly presentation in his book \textit{Communicating and Mobile Systems: the Pi-Calculus} in 1999 \cite{milner1999communicating} .

Was he going to be satisfied now?

\section{The Quest Continues}

For all the beauty and the success of the $\pi$-calculus, Robin realized that it was not \emph{the} $\lambda$-calculus of concurrency. There were too many possible variations and 
alternatives.\footnote{This is indeed an instance of one of the deepest issues of our subject: the `Next 700 \ldots' syndrome.}

With immense drive, energy and conviction Robin rewrote the script again.
Now he sought, rather than a single unique, comprehensive \emph{calculus}, to find the canonical, comprehensive structure at the \emph{metalevel} (cf. rewriting systems).

This led to the work on \emph{action calculi} and \emph{control structures} \cite{milner1993action,mifsud1995control}. This uncovered much fascinating static structure, but for a while seemed stalled on the issue of capturing the behavioural dynamics of systems at this level of generality.

Once again, Robin had other preoccupations for a time, as Head of Department of Cambridge.
What next?

\section{Bigraphs}


Robin's last phase of work was on bigraphs, culminating in his last book, published in 2009, on
\textit{The Space and Motion of Communicating Agents} \cite{milner2009space}.

Once again, Robin recast the paradigm. Here are some salient features:

\begin{itemize}
\item Two kinds of linking structure are recognized, which are orthogonal to a surprising degree: one the kind of linking by naming introduced in the $\pi$-calculus, the other a \emph{spatial structure}, as introduced by Cardelli and Gordon in the Ambient Calculus, and subsequently developed extensively as a sub-paradigm.
\item The \emph{static structure} is analyzed in terms of symmetric monoidal categories, building on the earlier work on action structures.
\item There is a remarkable treatment of the dynamics, solving the problems which had stymied the earlier work, using ideas of \emph{relative pushouts} in a highly original way --- by no means ``off-the-shelf'' category theory!
\item A natural and compelling graphical formalism.
\item A major push towards tool support and applications, in Ubiquitous Computing and Systems Biology, led, encouraged and guided by Robin.
\end{itemize}

\noindent How important will this new paradigm be? It is too soon to tell; but I wouldn't bet against it \ldots

\section{Some lessons we can learn from Robin}

\begin{itemize}
\item \emph{No Stone Tablets! No Disciples!}

Having created a paradigm, he recast it and made it new, not once, but in three major phases. Always with a purpose, moving the subject to a higher level.

\item \emph{Follow through!}

Four major books on concurrency: 1980, 1989, 1999, 2009. Led, inspired and encouraged a broad body of work, from theory to tool support and applications.

\item \emph{Be open to the work of others, and learn from them}. 

(Although in Robin's hands, the ideas were usually transformed in some way!)

Key examples include:
Structural Operational Semantics (Gordon Plotkin), bisimulation (David Park), structural congruence (G\'erard Berry and G\'erard Boudol), monoidal categories of processes (Jos\'e Meseguer and Ugo Montanari), spatial structure (Luca Cardelli and Andy Gordon).

\item Use \emph{the right mathematics} to realize your scientific vision; don't tailor your approach to suit some preconceived mathematics you happen to like.

The mathematical structures used by Robin in various phases of his work included domains, labelled transition systems and SOS, categories, and more.

\item \emph{Think it through}.

Whenever anyone raised a question concerning why such-and-such feature of one of his calculi was the way it was, or suggested some possible alternative, it became clear that Robin had considered all of these issues, carefully and deeply.

Robin was not given to lightning-fast rejoinders or intellectual pyrotechnics. Intellectually, he had
the speed of the long-distance runner.\footnote{Yoram Hirshfeld, who spent some time working closely with Robin, once remarked something on these lines, which has stayed with me.}
His ideas, too, have shown their staying power.
\end{itemize}

\noindent There are many more \ldots


\section{A final word}

Robin was \emph{both} a great scientist, \emph{and} a great human being.
Our community has been fortunate indeed to have someone so inspiring in human as well as intellectual terms.


%
\begin{center}
{\Large Thank You Robin.}
\end{center}

\bibliographystyle{entcs}

\bibliography{rmbib}
\end{document}